%% file: 0.0_CRP.tex
\documentclass[10pt,letterpaper]{article}
\usepackage[top=0.85in,left=1.00in,footskip=0.75in]{geometry}
\usepackage[utf8]{inputenc}
\usepackage{amsmath,amssymb}
\usepackage{cite}
\usepackage[round, sort&compress, numbers]{natbib}

\usepackage{changepage}

\usepackage{color}
\usepackage[table,xcdraw,dvipsnames]{xcolor}
\definecolor{cuteBlue}{rgb}{0.258, 0.387, 0.574}

\usepackage{nameref}
\usepackage[colorlinks=true, urlcolor=cuteBlue, citecolor=black, linkcolor=black]{hyperref}

\usepackage{microtype}
\DisableLigatures[f]{encoding = *, family = *}

\usepackage[aboveskip=1pt,labelfont=bf,labelsep=period,justification=raggedright,singlelinecheck=off]{caption}

\makeatletter
\renewcommand{\@biblabel}[1]{\quad#1.}
\makeatother

\usepackage{xparse}

\DeclareDocumentCommand \eref{oooo} {\IfNoValueTF{#2}{Eq.~\ref{#1}}{\IfNoValueTF{#3}{Eqs.~\ref{#1} and \ref{#2}}{\IfNoValueTF{#4}{Eqs.~\ref{#1}-\ref{#3}}{Eqs.~\ref{#1}-\ref{#4}}}}}

\DeclareDocumentCommand \fref{ooo} {\IfNoValueTF{#2}{Fig.~\ref{#1}}{\IfNoValueTF{#3}{Figs.~\ref{#1} and \ref{#2}}{Figs.~\ref{#1}-\ref{#3}}}}

\newcommand{\letter}[1]{(#1)} 
\newcommand{\letterParen}[1]{(#1)} 

\usepackage{graphicx}
\usepackage{multirow}

\usepackage{units, xfrac}

\begin{document}
	
	\include{1.1_Text}
	\include{2.1_Text_References} 
	
\end{document}

%% file: 1.1_Text.tex
\begin{flushleft} 
	{\Large \textbf\newline{Theoretical Analysis of Inducer and Operator Binding for Cyclic-AMP Receptor Protein Mutants}}
	\newline
	\\
	\textbf{Tal Einav$^{1}$, Julia Duque$^{2}$, Rob Phillips$^{1,3,4,*}$}
	\\
	$^1$Department of Physics, California Institute of Technology,
	Pasadena, CA, 91125, USA\\
	$^2$Department of Physics and Astronomy, London Centre for Nanotechnology,
	University College London, London, WC1H 0AH, United Kingdom\\
	$^3$Division of Biology and Biological Engineering, California Institute of Technology,
	Pasadena, CA, 91125, USA\\
	$^4$Department of Applied Physics, California Institute of Technology,
	Pasadena, CA, 91125, USA
	\\
	*Corresponding author: phillips@pboc.caltech.edu
\end{flushleft}

\section*{Abstract}

Allosteric transcription factors undergo binding events both at their inducer
binding sites as well as at distinct DNA binding domains, and it is often
difficult to disentangle the structural and functional consequences of these two
classes of interactions. In this work, we compare the ability of two statistical
mechanical models – the Monod-Wyman-Changeux (MWC) and the
Koshland-Némethy-Filmer (KNF) models of protein conformational change – to
characterize the multi-step activation mechanism of the broadly acting
cyclic-AMP receptor protein (CRP). We first consider the allosteric transition
resulting from cyclic-AMP binding to CRP, then analyze how CRP binds to its
operator, and finally investigate the ability of CRP to activate gene
expression. In light of these models, we examine data from a beautiful recent
experiment that created a single-chain version of the CRP homodimer, thereby
enabling each subunit to be mutated separately. Using this construct, six
mutants were created using all possible combinations of the wild type subunit, a
D53H mutant subunit, and an S62F mutant subunit. We demonstrate that both the
MWC and KNF models can explain the behavior of all six mutants using a small,
self-consistent set of parameters. In comparing the results, we find that the
MWC model slightly outperforms the KNF model in the quality of its fits, but
more importantly the parameters inferred by the MWC model are more in line with
structural knowledge of CRP. In addition, we discuss how the conceptual
framework developed here for CRP enables us to not merely analyze data
retrospectively, but has the predictive power to determine how combinations of
mutations will interact, how double mutants will behave, and how each construct
would regulate gene expression.

\section*{Introduction}	

Cyclic-AMP receptor protein (CRP; also known as the catabolite receptor protein,
CAP) is an allosteric transcription factor that regulates over 100 genes in
\textit{Escherichia coli} \cite{Martinez-Antonio2003, You2013, Vilar2013,
	Gama-Castro2016}. Upon binding to cyclic-AMP (cAMP), the homodimeric CRP
undergoes a conformational change whereby two alpha helices reorient to open a
DNA binding domain \cite{Popovych2009}, allowing CRP to bind to DNA and affect
transcription \cite{Hudson1990, Kuhlman2007, Kochanowski2017}. While much is
known about the molecular details of CRP and how different mutations modify its
functionality \cite{Youn2008, Sharma2009}, each new CRP mutant is routinely
analyzed in isolation using phenomenological models. We argue that given the
hard-won structural insights into the conformational changes of proteins like
CRP, it is important to test how well mechanistically motivated models of such
proteins can characterize the wealth of available data.

One of the difficulties inherent in understanding allosteric transcription
factors such as CRP lies in our inability to disentangle the numerous processes
involved in transcription, such as the binding of cAMP to CRP, of CRP to DNA,
and of transcription regulation as shown in \fref[figOverview]\letter{A}.
Measurements of gene expression depend upon all of these processes, and
mechanistic statements about the individual steps must necessarily be inferred
\cite{Garner1982, Fried1984, Takahashi1989, Bintu2005a}. To this end, \textit{in
	vitro} studies are beginning to probe each binding event separately, providing a
testbed to refine our understanding of both allostery and transcriptional
activation.

\begin{figure}[t]
	\centering \includegraphics{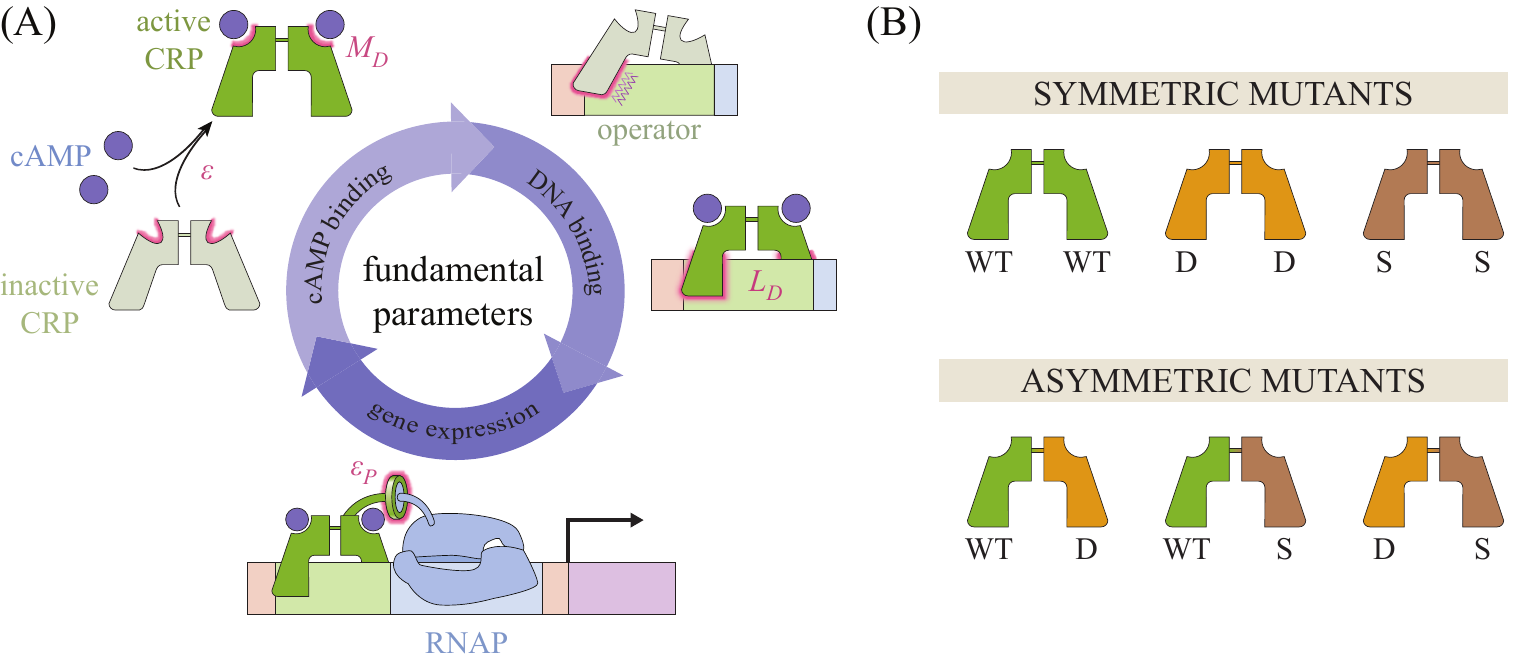}
	
	\caption{\textbf{Key parameters governing CRP function.} \letterParen{A} Within
	the MWC and KNF models, each CRP subunit can assume either an active or an
	inactive conformation with a free energy difference $\epsilon$ between the two
	states. cAMP can bind to CRP (with a dissociation constant $M_D^A$ in the
	active state and $M_D^I$ in the inactive state) and promotes the active state
	($M_D^A < M_D^I$ in the MWC model; $M_D^I \to \infty$ in the KNF model). Active
	CRP has a higher affinity for the operator ($L_D^A$) than the inactive state
	($L_D^I$). When CRP is bound to DNA, it promotes RNA polymerase binding through
	an interaction energy $\epsilon_P$, thereby enhancing gene expression.
	\letterParen{B} Lanfranco \textit{et al.}~constructed a single-chain CRP
	molecule whose two subunits could be mutated independently. They measured the
	cAMP and DNA binding affinity for CRP mutants comprised of wild type (WT), D
	(D53H), or S (S62F) subunits.} \label{figOverview}
\end{figure}

Our paper is inspired by a recent \textit{in vitro} study of CRP performed by
Lanfranco \textit{et al.}~who separately measured the two binding events of CRP,
first to cAMP and then to DNA \cite{Lanfranco2017}, providing an opportunity to
make a rigorous, quantitative analysis of the allosteric properties of CRP. To
this end, we explore two mechanistic frameworks for the allosteric transition of
CRP: the Monod-Wyman-Changeux (MWC) model, which posits that both CRP subunits
fluctuate concurrently between an active and inactive conformational state
\cite{Monod1965}, and the Koshland-Némethy-Filmer (KNF) model, which proposes
that each subunit must independently transition from an inactive to active state
upon ligand binding \cite{Koshland1966}. Although the MWC model provides a
marginally better characterization of the data, the two models offer different
interpretations of the general behavior of CRP. For example, the parameter set
we inferred for the KNF model predicts that CRP will bind equally well to DNA
regardless of whether it is bound to one or two cAMP molecules, while the MWC
model predicts that singly bound CRP will bind more tightly to DNA than unbound
and doubly bound CRP. Knowledge of the structure of CRP is in line with the MWC
predictions \cite{Heyduk1989, Sharma2009}, demonstrating that a model should not
be judged merely by its goodness of fit, but rather by the interpretation of its
fit parameters compared with the available knowledge of the system.

In addition to their consideration of the wild type protein, Lanfranco
\textit{et al.}~engineered a single-chain CRP molecule whose two subunits are
tethered together by an unstructured polypeptide linker. This construct enabled
them to mutate each subunit independently as shown in
\fref[figOverview]\letter{B}, providing a novel setting within which to analyze
the combinatorial effects of mutations. Specifically, they took three distinct
CRP subunits -- the wild type subunit and the well characterized mutations D53H
and S62F originally chosen to perturb the transcription factor's cAMP binding
domain \cite{Lin2002a, Dai2004} -- and linked them together in every possible
combination to create six CRP mutants.

The effects of mutations are often difficult to interpret, and indeed the
results from Lanfranco \textit{et al.}~showed no clear pattern. The behavior of
each mutant was analyzed independently by fitting its binding curve to a second
order polynomial, but this analysis was unable to make use of the fact that the
six CRP mutants are linked through their subunit compositions
\cite{Lanfranco2017}. In this work, we aim to close that gap by constructing a
quantitative framework that can describe the full suite of CRP data by utilizing
the subunit composition of CRP.

This concrete link between the composition and behavior of CRP mutants raises
the question of whether the response of a mutant can be predicted based on the
behavior of closely related mutants. For example, given sufficient data of CRP
with two wild type (WT) subunits and of CRP with two D53H subunits, can we
predict how a CRP comprised of one wild type and one D53H subunit will behave?
More generally, can the behavior of the symmetric mutants (top row of
\fref[figOverview]\letter{B}) predict the behavior of the asymmetric mutants
(bottom row of \fref[figOverview]\letter{B})? We demonstrate that both the MWC
and KNF models can generate such predictions. These results suggest a way to
harness the combinatorial complexity of oligomeric proteins and present a
possible step towards systematically probing the space of mutations.

We end by exploring the physiological impact of these CRP mutants by considering
how they would promote gene expression \textit{in vivo}. Because CRP is a global
activator, its activity within the cell is tightly regulated by enzymes that
produce, degrade, and actively transport cAMP. We discuss how these processes
can either be modeled theoretically or excised experimentally and calibrate our
resulting framework for transcription using gene expression measurements for
wild type CRP \cite{Kuhlman2007}. In this manner, we find a small,
self-consistent set of parameters able to characterize each step of CRP
activation shown in \fref[figOverview]\letter{A}.

The remainder of this paper is organized as follows. First, we characterize the
interaction between cAMP and CRP for the six CRP mutants created by Lanfranco
\textit{et al.}~and quantify the key parameters governing this behavior. Next,
we analyze the interaction between CRP and DNA from the perspectives of the MWC
and KNF models and discuss how the interpretations of the parameters inferred by
the MWC model are more in line with structural knowledge of the system. Finally,
we consider how CRP enhances gene expression and extend the results from
Lanfranco \textit{et al.}~to predict the activation profiles of the CRP mutants
within a cellular environment.

\section*{Results and Discussion}

\subsection*{The Interaction between CRP and cAMP}

In this section, we examine the cAMP-CRP binding process through the lenses of
the MWC and KNF models. We find that both models can characterize data from a
suite of CRP mutants using a compact set of parameters, thereby highlighting how
each mutant's behavior is tied to its subunit composition.

\subsubsection*{MWC Model}

We first formulate a description of cAMP-CRP binding using the MWC model, where
the two subunits of each CRP molecule fluctuate concurrently between an active
and inactive state. We define the free energy difference between the inactive
and active conformations to be $\epsilon$ per subunit, so that the total free
energy difference between inactive CRP and active CRP is $2\epsilon$. The
different conformations of CRP binding to cAMP and their corresponding Boltzmann
weights are shown in \fref[figCRPcAMPStatesWeights]\letter{A}. For each cAMP-CRP
dissociation constant $M_X^Y$, the subscript denotes which CRP subunit it
describes -- either the left ($L$) or right ($R$) subunit -- while the
superscript denotes the active ($A$) or inactive ($I$) state of CRP. Given a
cAMP concentration $[M]$, the fraction of occupied cAMP binding sites is given
by
\begin{equation} \label{eqCRPBindingFractionalOccupancy}
	\text{fractional CRP occupancy}([M]) = \frac{\frac{1}{2} \left( \frac{[M]}{M_L^A} + \frac{[M]}{M_R^A} \right) + \frac{[M]}{M_L^A} \frac{[M]}{M_R^A} + \frac{1}{2} e^{-2\beta\epsilon} \left( \frac{[M]}{M_L^I} + \frac{[M]}{M_R^I} \right) + e^{-2\beta\epsilon} \frac{[M]}{M_L^I} \frac{[M]}{M_R^I}}{\left( 1 + \frac{[M]}{M_L^A} \right) \left( 1 + \frac{[M]}{M_R^A} \right) + e^{-2\beta\epsilon} \left( 1 + \frac{[M]}{M_L^I} \right) \left( 1 + \frac{[M]}{M_R^I} \right)}.
\end{equation}
Here, the fractional occupancy of CRP bound to zero, one, or two cAMP equals 0,
$\nicefrac{1}{2}$, and 1, respectively. Experimentally, the fractional occupancy
was measured using ANS fluorescence which utilizes a fluorescent probe triggered
by the conformational change of cAMP binding to CRP \cite{Lanfranco2017}. We
note that this measurement was carried out \textit{in vitro} in the absence of
DNA.

\begin{figure}[t]
	\centering \includegraphics{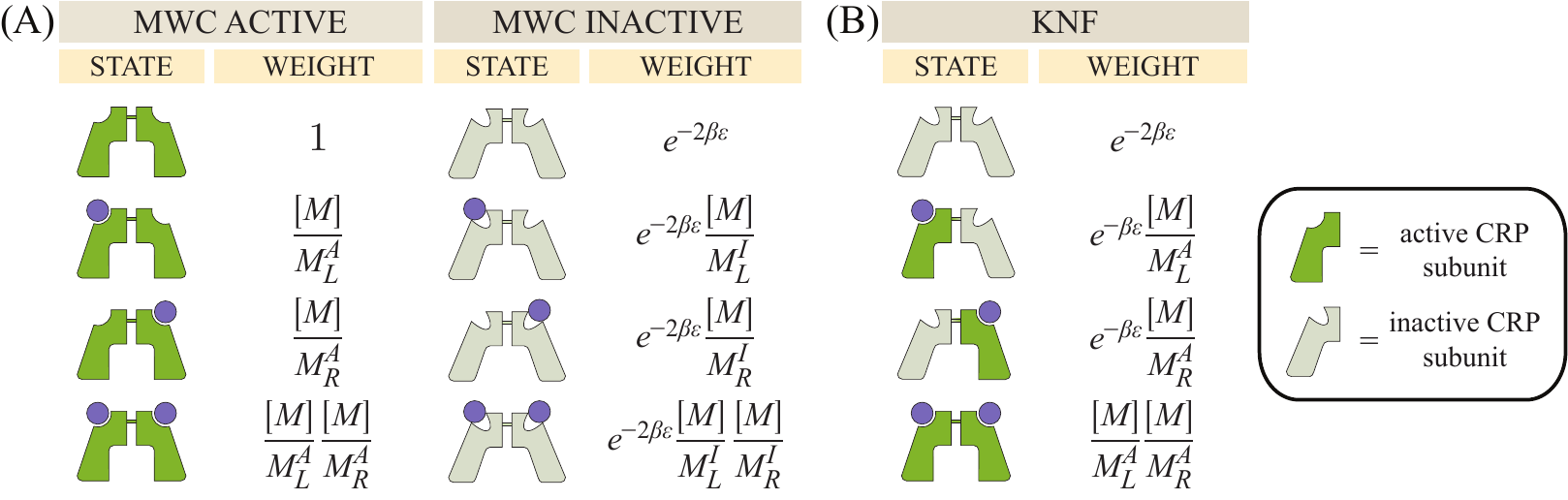}
	
	\caption{\textbf{Macroscopic states and Boltzmann weights for cAMP binding to
		CRP.} \letterParen{A} Within the MWC model, cAMP (purple circles) may bind to a
	CRP subunit in either the active (dark green) or inactive (light green) state.
	$M_L^A$ and $M_L^I$ represent the dissociation constants of the left subunit in
	the active and inactive states, respectively, while $M_R^A$ and $M_R^I$
	represent the analogous dissociation constants for the right subunit. $[M]$
	denotes the concentration of cAMP and $\epsilon$ represents the free energy
	difference between each subunit's inactive and active states. \letterParen{B}
	The KNF model assumes that the two CRP subunits are inactive when unbound to
	cAMP and transition to the active state immediately upon binding to cAMP. The
	parameters have the same meaning as in the MWC model, but states where one
	subunit is active while the other is inactive are allowed.}
\label{figCRPcAMPStatesWeights}
\end{figure}

Lanfranco \textit{et al.}~considered CRP subunits with either the D53H or S62F
point mutations (hereafter denoted by D and S, respectively), with the D subunit
binding more strongly to cAMP than the wild type while the S subunit binds more
weakly as shown in \fref[figCRPcAMPligandTitration]\letter{A}. While we could
characterize the dose-response curves of each CRP mutant independently -- for
example, by using \eref[eqCRPBindingFractionalOccupancy] to extract a set of
parameters for each mutant -- such an analysis lacks a direct connection between
the subunit composition and the corresponding binding behavior. Instead, we
assume that the cAMP binding affinity for each subunit should be uniquely
dictated by that subunit's identity as either the WT, D, or S subunit. To that
end, we represent the fractional occupancy of $\text{CRP}_\text{D/WT}$ using
\eref[eqCRPBindingFractionalOccupancy] with one D subunit ($M_L^A =
M_\text{D}^A$, $M_L^I = M_\text{D}^I$) and one WT subunit ($M_R^A =
M_\text{WT}^A$, $M_R^I = M_\text{WT}^I$). The equations for the remaining CRP
mutants follow analogously, tying the behavior of each mutant to its subunit
composition.

\begin{figure}[t]
	\centering \includegraphics{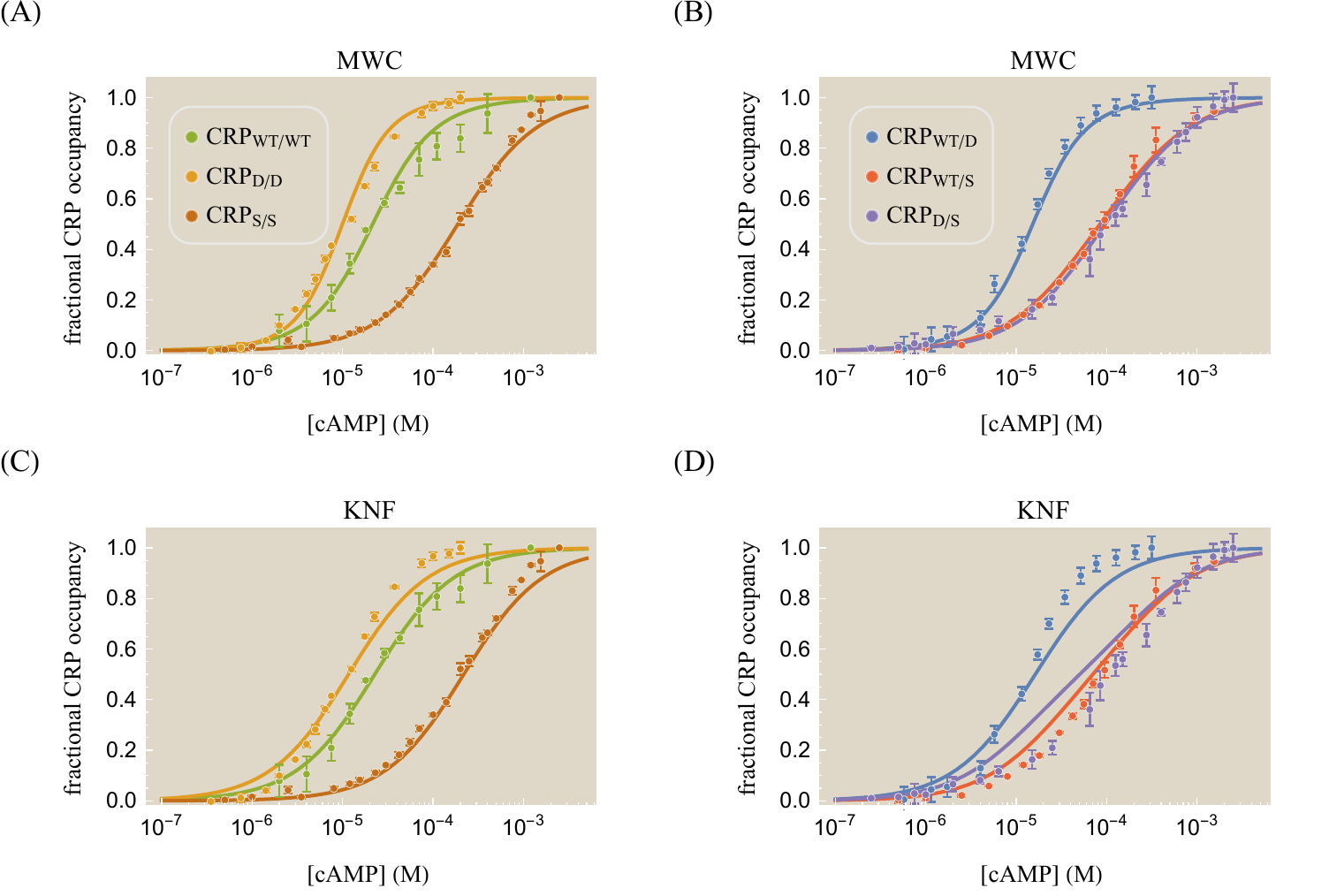}
	
	\caption{\textbf{cAMP binding curves for different CRP mutants.} In addition to
	the wild type CRP subunit (denoted WT), the mutation D53H (denoted D) and the
	mutation S62F (denoted S) can be applied to either subunit as indicated by the
	subscripts in the legend. \letterParen{A} Curves were characterized using the
	MWC model, \eref[eqCRPBindingFractionalOccupancy]. The D subunit increases
	CRP's affinity for cAMP while the S subunit decreases this affinity.
	\letterParen{B} Asymmetrically mutating the two subunits results in distinct
	cAMP binding curves. The data for the WT/D mutant lies between the WT/WT and
	D/D data in Panel \letter{A}, and analogous statements apply for the WT/S and
	D/S mutants. \letterParen{C} The symmetric mutants and \letterParen{D} the
	asymmetric mutants can also be analyzed using the KNF model,
	\eref[eqCRPBindingFractionalOccupancyKNF], resulting in curves that are similar
	to those found by the MWC model. The (corrected) sample standard deviation
	$\sqrt{\frac{1}{n-1} \sum_{j=1}^n (y_\text{theory}^{(j)} -
		y_\text{data}^{(j)})^2}$ equals $0.03$ for the MWC model and $0.06$ for the KNF
	model, and the best-fit parameters for both models are given in
	Table~\ref{tableCRPparamaters}.} \label{figCRPcAMPligandTitration}
\end{figure}

\begin{table}[t]
	\begin{center}
		\caption{\textbf{Parameters for cAMP binding to CRP.} The data in
	\fref[figCRPcAMPligandTitration] can be characterized using a single set of
	dissociation constants for the WT, D, and S subunits whose values and standard
	errors are shown. The left column corresponds to the MWC parameters given in
	\eref[eqCRPBindingFractionalOccupancy] while the right column corresponds to
	the KNF model given by \eref[eqCRPBindingFractionalOccupancyKNF]. For both
	models, the data only constrains the parameter combination $\tilde{M}_X^A =
	M_X^A e^{-\beta\epsilon}$. The effective dissociation constant of the S
	subunit in the MWC model can only be bounded from below as
	$\tilde{M}_\text{S}^A \ge 1000 \times 10^{-6}~\text{M}$.}
		
		\begin{tabular}{cc|cc}
			\hline
			\textbf{MWC Parameter} 			& \textbf{Best-Fit Value} 				& \textbf{KNF Parameter} 	& \textbf{Best-Fit Value}			\\ \rowcolor[HTML]{EFEFEF}
			\hline \hline
			$\tilde{M}_\text{WT}^A$, $M_\text{WT}^I$& $\{25 \pm 1, 40 \pm 3 \} \times 10^{-6}~\text{M}$	& $\tilde{M}_\text{WT}^A$	& $(20 \pm 2) \times 10^{-6}~\text{M}$		\\ 
			$\tilde{M}_\text{D}^A$, $M_\text{D}^I$ 	& $\{10 \pm 1, 50 \pm 5 \} \times 10^{-6}~\text{M}$	& $\tilde{M}_\text{D}^A$	& $(10 \pm 1) \times 10^{-6}~\text{M}$ 		\\ \rowcolor[HTML]{EFEFEF}
			$\tilde{M}_\text{S}^A$, $M_\text{S}^I$	& $\{\ge 1000, 200 \pm 10 \} \times 10^{-6}~\text{M}$	& $\tilde{M}_\text{S}^A$	& $(220 \pm 10) \times 10^{-6}~\text{M}$ 	\\
			\hline
			\label{tableCRPparamaters}
		\end{tabular}
	\end{center}
\end{table}

Many studies have been conducted upon cAMP-CRP binding using a broad range of
methods including equilibrium dialysis, fluorescence assays, and protease
digestion, but the resulting parameters inferred from these experiments have
varied widely \cite{Lin2002a}. Estimates on the free energy difference
$2\epsilon$ between inactive and active CRP range from approximately $-20\,k_B
T$ to $-10\,k_B T$ \cite{Cheng1998, Sharma2009, Gunasekara2015} while apparent
dissociation between cAMP and $\text{CRP}_\text{WT/WT}$ range from
$10^{-6}\mbox{-}10^{-3}\,\text{M}$ \cite{Emmer1970, Suh2002, Lin2002a}.

Part of the difficulty in pinning down these values stems from the fact that
degenerate parameter values reproduce equivalent binding curves. For instance,
it is known that in the absence of cAMP, the overwhelming majority of CRP
molecules will be in the inactive state ($1 \ll e^{-2 \beta \epsilon}$).
Intuitively, there will effectively be no active CRP molecules in the absence of
cAMP regardless of whether $2\epsilon = -10\,k_B T$ or $2\epsilon = -20\,k_B T$;
however, in the presence of saturating cAMP the latter case will require a
larger affinity between active CRP and cAMP (smaller $M_L^A$ and $M_R^A$) to
compensate for this greater free energy difference. The same cAMP-CRP binding
curves can even be produced for an arbitrarily large and negative free energy
difference ($\epsilon \to - \infty$) provided that the dissociation constants
scale appropriately. This scaling can be determined by fixing the ratio of
doubly-cAMP bound CRP in the active state ($\frac{[M]}{M_L^A}
\frac{[M]}{M_L^A}$) and inactive state ($e^{-2 \beta \epsilon} \frac{[M]}{M_L^I}
\frac{[M]}{M_R^I}$). For the CRP system this requires that $M_L^I$ and $M_R^I$
do not depend on $\epsilon$ while $M_L^A, M_R^A \propto e^{\beta \epsilon}$.
Mathematically, in the limit of a large and negative $\epsilon$ the cAMP
occupancy becomes
\begin{align} \label{eqReducedFractionalOccupancy}
	\text{fractional CRP occupancy}([M]) &= \frac{\frac{1}{2} e^{\beta\epsilon} \left( \frac{[M]}{\tilde{M}_L^A} + \frac{[M]}{\tilde{M}_R^A} \right) + \frac{[M]}{\tilde{M}_L^A} \frac{[M]}{\tilde{M}_R^A} + \frac{1}{2} \left( \frac{[M]}{M_L^I} + \frac{[M]}{M_R^I} \right) + \frac{[M]}{M_L^I} \frac{[M]}{M_R^I}}{\left( e^{\beta\epsilon} + \frac{[M]}{\tilde{M}_L^A} \right) \left( e^{\beta\epsilon} + \frac{[M]}{\tilde{M}_R^A} \right) + \left( 1 + \frac{[M]}{M_L^I} \right) \left( 1 + \frac{[M]}{M_R^I} \right)} \nonumber\\
	&\approx \frac{\frac{[M]}{\tilde{M}_L^A} \frac{[M]}{\tilde{M}_R^A} + \frac{1}{2} \left( \frac{[M]}{M_L^I} + \frac{[M]}{M_R^I} \right) + \frac{[M]}{M_L^I} \frac{[M]}{M_R^I}}{ \frac{[M]}{\tilde{M}_L^A} \frac{[M]}{\tilde{M}_R^A} + \left( 1 + \frac{[M]}{M_L^I} \right) \left( 1 + \frac{[M]}{M_R^I} \right)},
\end{align}
where in the first equality we multiplied the numerator and denominator of \eref[eqCRPBindingFractionalOccupancy] by $e^{2 \beta \epsilon}$ and defined the effective dissociation constants
\begin{equation} \label{eqMLAEffective}
	\tilde{M}_L^A = e^{-\beta \epsilon} M_L^A,
\end{equation}
and
\begin{equation} \label{eqMRAEffective}
	\tilde{M}_R^A = e^{-\beta \epsilon} M_R^A,
\end{equation}
while in the latter equality of \eref[eqReducedFractionalOccupancy] we neglected
all of the terms multiplied by the small quantity $e^{\beta\epsilon}$ (which is
equivalent to taking the zeroth order Taylor series about $e^{-\beta \epsilon}
\approx 0$). In Supporting Information Section
A, we demonstrate how the $\epsilon$
parameter may be shifted arbitrarily without altering the cAMP-CRP binding
curves provided \eref[eqMLAEffective][eqMRAEffective] hold and that $2\epsilon
\lesssim -3\,k_B T$ (above which the approximation $1 \ll e^{-2 \beta \epsilon}$
breaks down). This last assumption is well justified, since the overwhelming
majority of CRP molecules assume the inactive conformation in the absence of
cAMP \cite{Sharma2009}. Therefore, the maximum information that can be extracted
from the data includes the inactive CRP-cAMP dissociation constants $M_L^I$ and
$M_R^I$ and the effective dissociation constants $\tilde{M}_L^A$ and
$\tilde{M}_R^A$. Lastly, we note that if the D and S mutations alter the free
energy $\epsilon$, that effect will be absorbed into the effective dissociation
constants.

Using \eref[eqReducedFractionalOccupancy], we can extract the set of effective
dissociation constants for the WT, D, and S subunits that determine the behavior
of all six CRP mutants. The resulting parameters (shown in
Table~\ref{tableCRPparamaters}) give rise to the cAMP-CRP binding curves in
\fref[figCRPcAMPligandTitration]\letter{A} and \letter{B}. In Supporting
Information Section~B, we demonstrate that the symmetric
CRP mutants in \fref[figCRPcAMPligandTitration]\letter{A} provide sufficient
information to predict the behavior of the asymmetric mutants in
\fref[figCRPcAMPligandTitration]\letter{B}. We further show that fitting each
CRP data set individually to the MWC or KNF models without constraining the WT,
D, and S subunits to a single unified set of dissociation constants results in
only a marginal improvement over the constrained fitting. Finally, we analyze
the slope of each cAMP binding response and explain why they are nearly
identical for the six CRP mutants. Supporting Information
Section~C investigates the effects of the double mutation
D+S on a single subunit.

\subsubsection*{KNF Model}

We now turn to a KNF analysis of CRP, where the two subunits are individually
inactive when not bound to cAMP and become active upon binding as shown in
\fref[figCRPcAMPStatesWeights]\letter{B}. Some studies have claimed that cAMP
binding to one CRP subunit does not affect the state of the other subunit, in
support of the KNF model \cite{Popovych2006}. Other studies, meanwhile, have
reported that a fraction of CRP molecules are active even in the absence of
cAMP, thereby favoring an MWC interpretation \cite{Youn2008}. It is not yet
known whether either model can accurately represent the system. To that end, we
explore some of the consequences of a KNF interpretation of CRP.

Using the statistical mechanical states of the system in
\fref[figCRPcAMPStatesWeights]\letter{B}, the occupancy of CRP is
given by
\begin{equation} \label{eqKNFfractionalOccupancy}
\text{fractional CRP occupancy}([M]) = \frac{\frac{e^{-\beta\epsilon}}{2} \left( \frac{[M]}{M_L^A} + \frac{[M]}{M_R^A} \right) + \frac{[M]}{M_L^A} \frac{[M]}{M_R^A}}{\left( e^{-\beta\epsilon} + \frac{[M]}{M_L^A} \right) \left( e^{-\beta\epsilon} + \frac{[M]}{M_R^A} \right)}.
\end{equation}
Note that by multiplying the numerator and denominator by $e^{2\beta\epsilon}$
and defining the same effective dissociation constants
(\eref[eqMLAEffective][eqMRAEffective]) as for the MWC model, we can eliminate the
free energy difference $\epsilon$ to obtain the form
\begin{equation} \label{eqCRPBindingFractionalOccupancyKNF}
\text{fractional CRP occupancy}([M]) = \frac{\frac{1}{2} \left( \frac{[M]}{\tilde{M}_L^A} + \frac{[M]}{\tilde{M}_R^A} \right) + \frac{[M]}{\tilde{M}_L^A} \frac{[M]}{\tilde{M}_R^A}}{\left( 1 + \frac{[M]}{\tilde{M}_L^A} \right) \left( 1 + \frac{[M]}{\tilde{M}_R^A} \right)}.
\end{equation}
This simplification occurs because within the KNF model, a CRP monomer only
switches from the inactive to active state upon cAMP binding. As a result, the
free energy of cAMP binding to CRP and the free energy of the CRP undergoing its
inactive-to-active state conformational always occur concurrently and may be
combined into the effective dissociation constants $\tilde{M}_L^A$ and
$\tilde{M}_R^A$. As shown in \fref[figCRPcAMPligandTitration]\letter{C} and
\fref[figCRPcAMPligandTitration]\letter{D}, the KNF model can characterize the
six mutant CRP binding curves, albeit with a larger sample standard deviation than
the MWC model. While the KNF model has fewer parameters, the cost of this
simplicity is manifest in its slightly poorer fits.

\subsection*{The Interaction between CRP and DNA}

We now turn to the second binding interaction experienced by CRP, namely, that
between CRP and DNA within the MWC and KNF models. 

\subsubsection*{MWC Model}

Consider a concentration $[L]$ of CRP whose subunits either assume an active
state (where they tightly bind to DNA with a dissociation constant $L_A$) or in
an inactive state (characterized by weaker DNA binding with dissociation
constant $L_I$ satisfying $L_I > L_A$). The states and weights of this system
within the MWC model are shown in \fref[figCRPStatesWeightsMWC]\letter{A}.

\begin{figure}[t]
	\centering \includegraphics{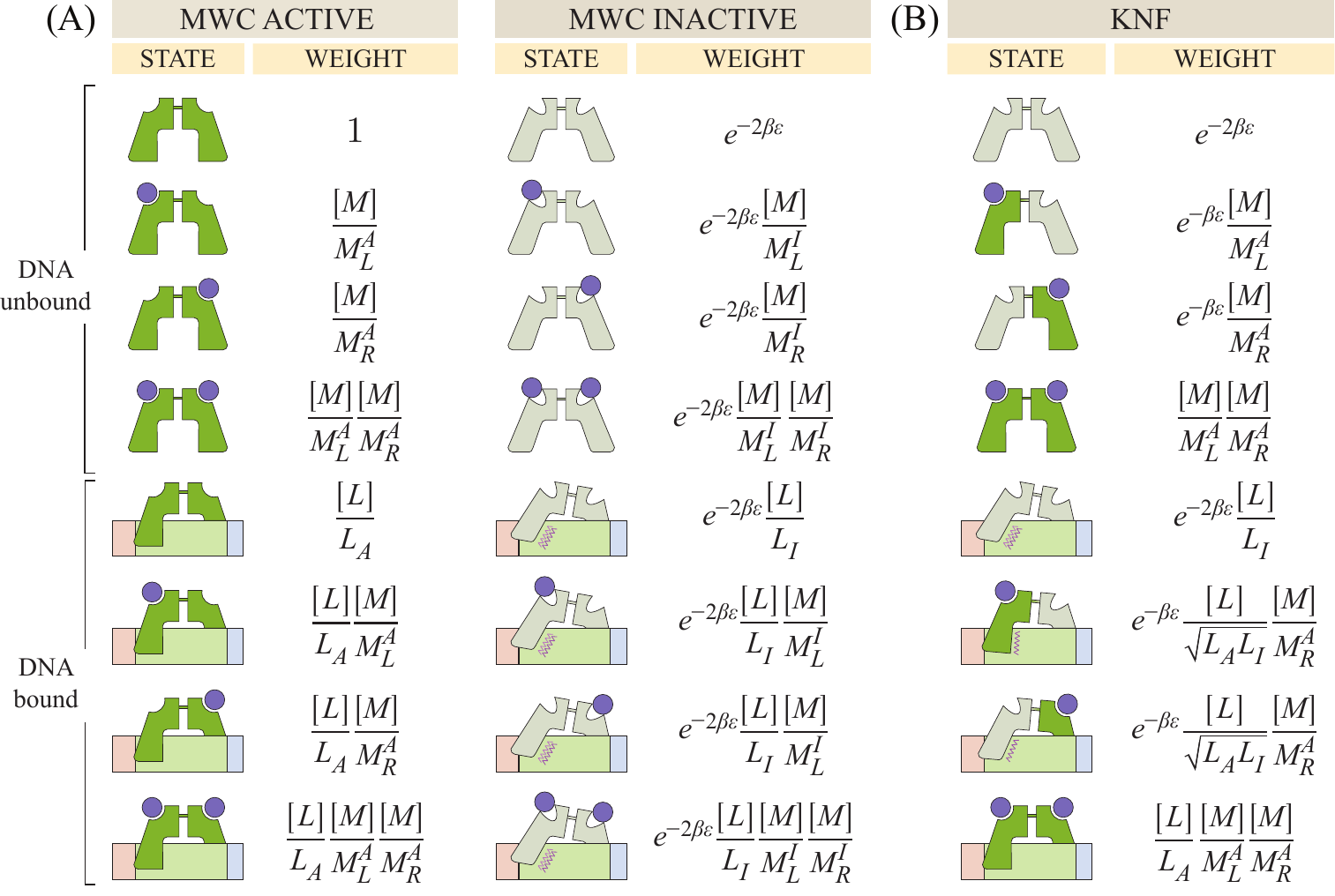}
	
	\caption{\textbf{States and weights for CRP binding to DNA.} \letterParen{A}
	The DNA unbound states from \fref[figCRPcAMPStatesWeights] together with the
	DNA bound states. The Boltzmann weight of each DNA bound state is proportional
	to the concentration $[L]$ of CRP and inversely proportional to the CRP-DNA
	dissociation constants $L_A$ or $L_I$ for the active and inactive states,
	respectively. \letterParen{B} In the KNF model, these same parameters apply to
	each CRP subunit which can be independently active or inactive.}
\label{figCRPStatesWeightsMWC}
\end{figure}

Lanfranco \textit{et al.}~fluorescently labeled a short, 32 bp DNA sequence
which binds to CRP. Using a spectrometer, they measured the anisotropy of this
fluorescence when different concentrations of CRP and cAMP were added \textit{in
	vitro} \cite{Lanfranco2017}. The data are shown in
\fref[figCRPDNAbindingMWCSimplest]\letter{A} for $\text{CRP}_{\text{D/S}}$. When
CRP binds, it slows the random tumbling of the DNA so that over very short time
scales the fluorescence is oriented along a particular axis, resulting in a
larger anisotropy readout. Unbound DNA is defined as having $\text{anisotropy} =
1$ while DNA-bound CRP with 0, 1, or 2 bound cAMP have higher anisotropies of
$1+r_0$, $1+r_1$, and $1+r_2$, respectively. Thus, the total anisotropy within
the model is given by the weighted sum of each species \cite{Heyduk1990},
namely,
\begin{equation} \label{eqAnisotropyCRPbinding}
\text{anisotropy} = 1 + r_0 p_0 + r_1 p_1 + r_2 p_2.
\end{equation} 
Here, $p_0$, $p_1$, and $p_2$ represent the probability that DNA-bound CRP will
be bound to 0, 1, and 2 cAMP molecules, respectively. In this model, we have extended
the classic MWC model to allow each of these states to have a unique DNA binding
affinity. Using the effective dissociation constants
(\eref[eqMLAEffective][eqMRAEffective]) and neglecting all terms proportional to
the small quantity $e^{\beta\epsilon}$, we can write these probabilities as
\begin{equation} \label{eqCRP0cAMPboundMWC}
	p_0 = \frac{e^{2\beta\epsilon} \frac{[L]}{L_A} + \frac{[L]}{L_I}}{Z} \approx \frac{\frac{[L]}{L_I}}{Z},
\end{equation}
\begin{equation} \label{eqCRP1cAMPboundMWC}
	p_1 = \frac{e^{2\beta\epsilon} \frac{[L]}{L_A} \left( \frac{[M]}{M_L^A} + \frac{[M]}{M_R^A} \right) + \frac{[L]}{L_I} \left( \frac{[M]}{M_L^I} + \frac{[M]}{M_R^I} \right)}{Z} \approx \frac{\frac{[L]}{L_I} \left( \frac{[M]}{M_L^I} + \frac{[M]}{M_R^I} \right)}{Z},
\end{equation}
and
\begin{equation} \label{eqCRP2cAMPboundMWC}
	p_2 = \frac{e^{2\beta\epsilon} \frac{[L]}{L_A} \frac{[M]}{M_L^A} \frac{[M]}{M_R^A} + \frac{[L]}{L_I} \frac{[M]}{M_L^I} \frac{[M]}{M_R^I}}{Z} \approx \frac{\frac{[L]}{L_A} \frac{[M]}{\tilde{M}_L^A} \frac{[M]}{\tilde{M}_R^A} + \frac{[L]}{L_I} \frac{[M]}{M_L^I} \frac{[M]}{M_R^I}}{Z}
\end{equation}
with
\begin{align} \label{eqCRPDenominatorcAMPboundMWC}
	Z &= e^{2\beta\epsilon} \left(1+\frac{[L]}{L_A}\right)\left(1+\frac{[M]}{M_L^A}\right)\left(1+\frac{[M]}{M_R^A}\right) + \left(1+\frac{[L]}{L_I}\right) \left(1+\frac{[M]}{M_L^I}\right)\left(1+\frac{[M]}{M_R^I}\right) \nonumber \\
	&\approx \left(1+\frac{[L]}{L_A}\right) \frac{[M]}{\tilde{M}_L^A} \frac{[M]}{\tilde{M}_R^A} + \left(1+\frac{[L]}{L_I}\right) \left(1+\frac{[M]}{M_L^I}\right)\left(1+\frac{[M]}{M_R^I}\right).
\end{align}
In making these approximations, we have assumed the stricter conditions
$e^{2\beta\epsilon}\frac{L_I}{L_A} \ll 1$ and
$e^{2\beta\epsilon}\frac{L_I}{L_A} \frac{M_X^I}{\tilde{M}_X^A} \ll 1$ for
the WT, D, and S subunits, all of which are valid assumptions for this system
(see Supporting Information Section~A).

\begin{figure}[t]
	\centering \includegraphics{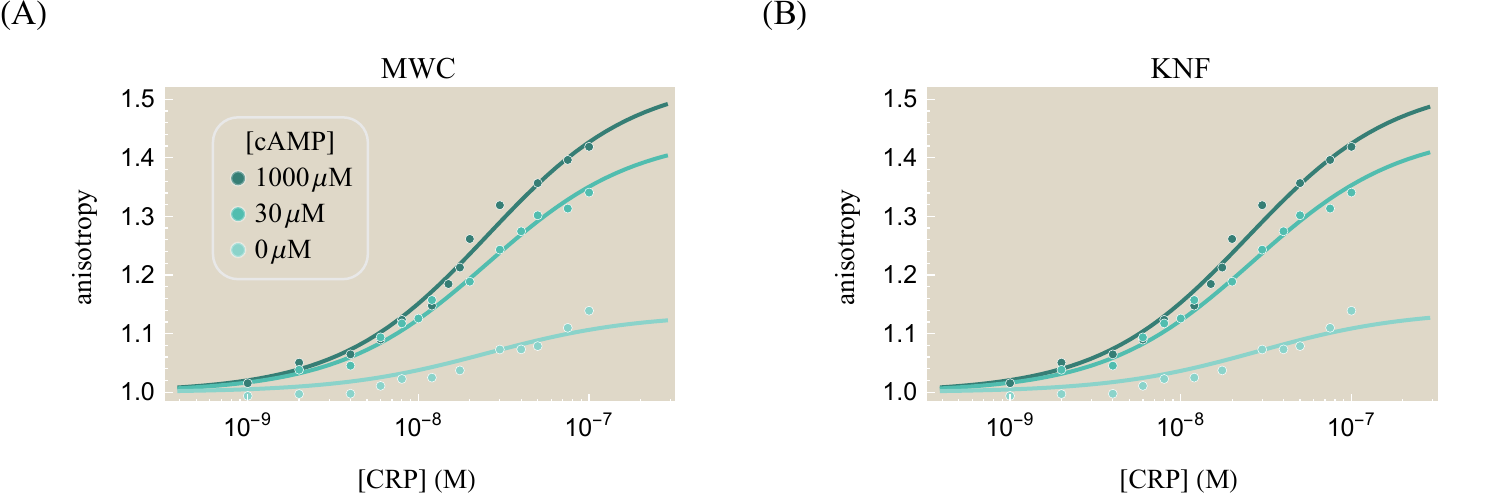}
	
	\caption{\textbf{The interaction between CRP and DNA.} \letterParen{A}
	Anisotropy of 32-bp fluorescein-labeled \textit{lac} promoter binding to
	$\text{CRP}_{\text{D/S}}$ at different concentrations of cAMP. An anisotropy of
	1 corresponds to unbound DNA while higher values imply that DNA is bound to
	CRP. \letterParen{B} This same data analyzed using the KNF model. In the
	presence of cAMP, more CRP subunits will be active, and hence there will be
	greater anisotropy for any given concentration of CRP. The sample standard
	deviation $\sqrt{\frac{1}{n-1} \sum_{j=1}^n (y_\text{theory}^{(j)} -
		y_\text{data}^{(j)})^2}$ is $0.01$ for both the MWC and KNF models, with the
	corresponding parameters given in Tables~\ref{tableCRPparamaters} and
	\ref{tableCRPDNABindingParamaters}.} \label{figCRPDNAbindingMWCSimplest}
\end{figure}

\fref[figCRPDNAbindingMWCSimplest]\letter{A} shows the resulting best-fit curves
for the anisotropy data, with the corresponding $\text{CRP}_{\text{D/S}}$ DNA
dissociation constants given in Table~\ref{tableCRPDNABindingParamaters}. Since
$1 + r_0 \approx 1$, unbound CRP binds poorly to DNA, in accordance with the
inactive state crystal structure whose DNA recognition helices are buried inside
the protein \cite{Sharma2009}. Additionally, the anisotropy $1 + r_1 = 1.7$ of
the DNA-CRP-cAMP complex is larger than that of both the unbound state and the
doubly bound state DNA-CRP-$\left(\text{cAMP}\right)_2$ with $1 + r_2 = 1.4$;
this suggests that CRP-$\left(\text{cAMP}\right)_2$ binds more weakly to DNA
that CRP-cAMP. Previous studies have confirmed this claim using multiple
experimental methods including proteolytic digestion by subtilisin, chemical
modification of Cys-178, and fluorescence measurements \cite{Heyduk1989,
	Pyles1996}, although other work has proposed an alternate explanation that above
millimolar cAMP concentrations CRP attains new states with highly unfavorable
DNA binding affinities \cite{Takahashi1989, Passner1997}. In Supporting
Information Section D, we extend the analysis of CRP-DNA
binding to the remaining CRP mutants.

\begin{table}[t]
	\begin{center}
		\caption{\textbf{Parameters for CRP binding to DNA.} The anisotropy data for
	$\text{CRP}_{\text{D/S}}$ characterized using \eref[eqAnisotropyCRPbinding],
	as shown in \fref[figCRPDNAbindingMWCSimplest]. Each value is given as a mean
	$\pm$ standard error. The uncertainty in $\tilde{M}_\text{S}^A$ parameter
	(shown in Table~\ref{tableCRPparamaters}) leads to a corresponding uncertainty
	in the active CRP dissociation constant $L_A$.}
		
		\begin{tabular}{cc|cc}
			\hline
			\textbf{MWC Parameter}								& \textbf{Best-Fit Value}				& \textbf{KNF Parameter}& \textbf{Best-Fit Value}				\\ \rowcolor[HTML]{EFEFEF}
			\hline \hline 
			$r_0$, $r_1$, $r_2$& $\{ 0.1, 0.8, 0.5 \} \pm 0.1$ 					& $r_0$, $r_1$, $r_2$ 		& $\{ 0.1, 0.5, 0.5 \} \pm 0.1$ 								\\ 
			$L_A$, $L_I$   										& $\{ \le 30, 30 \pm 10 \} \times 10^{-9}~\text{M}$	& $L_A$, $L_I$ 			& $\{ 20 \pm 10, 30 \pm 20 \} \times 10^{-9}~\text{M}$ \\
			\hline
			\label{tableCRPDNABindingParamaters}
		\end{tabular}
	\end{center}
\end{table}

\subsubsection*{KNF Model}

We now turn to the KNF model of DNA binding where each CRP subunit is inactive
when not bound to cAMP and active when bound to cAMP as shown in
\fref[figCRPStatesWeightsMWC]\letter{B}. As in the MWC model, $L_A$ and $L_I$
represent the dissociation constants between DNA and CRP in the active and
inactive states, respectively. The mixed state of DNA-bound CRP with one active
and one inactive subunit has a dissociation constant $\sqrt{L_A L_I}$ (see
Supporting Information Section E). The anisotropy
within the KNF model is given by \eref[eqAnisotropyCRPbinding] with
\begin{equation} \label{eqCRP0cAMPboundKNF}
p_0 = \frac{\frac{[L]}{L_I}}{Z},
\end{equation}
\begin{equation} \label{eqCRP1cAMPboundKNF}
p_1 = \frac{\frac{[L]}{\sqrt{L_A L_I}} \left( \frac{[M]}{\tilde{M}_L^A} + \frac{[M]}{\tilde{M}_R^A} \right)}{Z},
\end{equation}
and
\begin{equation} \label{eqCRP2cAMPboundKNF}
p_2 = \frac{\frac{[L]}{L_A} \frac{[M]}{\tilde{M}_L^A} \frac{[M]}{\tilde{M}_R^A}}{Z}
\end{equation}
representing the probabilities of the DNA-CRP, DNA-CRP-cAMP, and
DNA-CRP-$\left(\text{cAMP}\right)_2$ states, respectively, with
\begin{equation}
Z = \left( 1 + \frac{[M]}{\tilde{M}_L^A} \right) \left( 1 + \frac{[M]}{\tilde{M}_R^A} \right) + \frac{[L]}{L_I} \left( 1 + \frac{[M]}{\tilde{M}_L^A} \sqrt{\frac{L_I}{L_A}} \right) \left( 1 + \frac{[M]}{\tilde{M}_R^A} \sqrt{\frac{L_I}{L_A}} \right).
\end{equation}

\fref[figCRPDNAbindingMWCSimplest]\letter{B} shows the characterization of the
anisotropy data using the KNF model with the corresponding parameters given by
the right column of Table~\ref{tableCRPDNABindingParamaters}. Unlike the MWC
model, the best-fit KNF anisotropy parameters satisfy $r_1 \approx r_2$,
suggesting that CRP binds equally tightly to DNA independent of how many cAMP
are bound to it. Structural understanding of CRP support the MWC
characterization, as both unbound and doubly bound CRP have been reported to have a
significantly lower DNA-binding affinity than singly-cAMP bound CRP
\cite{Heyduk1989, Pyles1996}.

\subsection*{Implications of Mutations for \textit{in vivo} Systems}

CRP is a global transcriptional activator which governs many metabolic genes in
\textit{E. coli} \cite{Kochanowski2017}. It is interesting to consider how the
mutants characterized in the Lanfranco \textit{et al.}~experiments would behave
as transcriptional activators for \textit{in vivo} gene expression measurements.
In this section, we make predictions for how CRP will act \textit{in vivo} for
the different mutants characterized above. To focus our analysis, we use the MWC
model presented above to analyze gene expression measurements of a system where
CRP is the only transcription factor present, although it is straightforward to
generalize to more complex regulatory architectures or to apply the KNF
framework to this process \cite{Bintu2005}.

\subsubsection*{Simple Activation}

As above, consider a cell with cAMP concentration $[M]$ and CRP concentration
$[L]$ where the population of CRP is split between an active $[L_A]$ and an
inactive $[L_I]$ conformation. Suppose the cell has a concentration $[P]$ of RNA
polymerase (RNAP) which have a dissociation constant $P_D$ with a promoter of
interest. The thermodynamic states of the system are shown in
\fref[figPromoterStatesWeights], where the activator can bind to and recruit
RNAP via an interaction energy $\epsilon_{P,L_A}$ between active CRP and RNAP
with a weaker interaction $\epsilon_{P,L_I}$ between inactive CRP and RNAP.
Without these two interaction energies, the RNAP and CRP binding events would be
independent and there would be no activation.

\begin{figure}[t]
	\centering \includegraphics{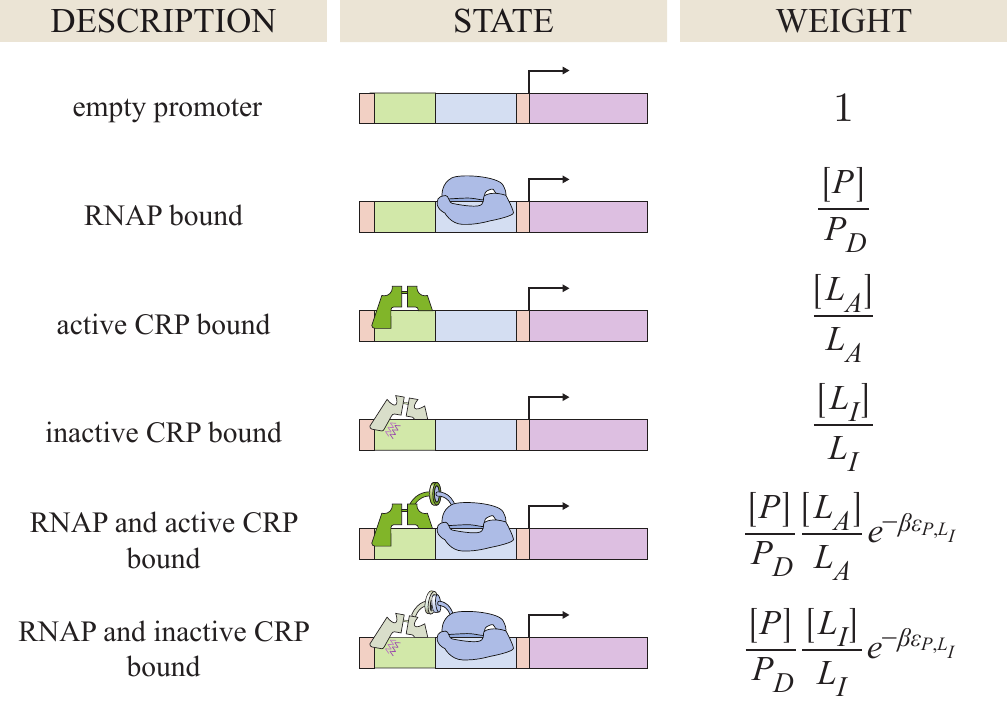}

	\caption{\textbf{States and weights for a simple activation motif.} Binding of
	RNAP (blue) to a promoter is facilitated by the binding of the activator CRP.
	Simultaneous binding of RNAP and CRP is facilitated by an interaction energy
	$\epsilon_{P,L_A}$ for active CRP (dark green) and $\epsilon_{P,L_I}$ for
	inactive CRP (light green). cAMP (not drawn) influences the concentration of
	active and inactive CRP as shown in \fref[figCRPStatesWeightsMWC].}
\label{figPromoterStatesWeights}
\end{figure}

We assume that gene expression is equal to the product of the RNAP transcription
rate $r_{\text{trans}}$ and the probability that RNAP is bound to the promoter
of interest, namely,
\begin{equation} \label{RNApBound}
\text{activity} = r_{\text{trans}} \frac{ \frac{[P]}{P_D} \left( 1 + \frac{[L_I]}{L_I}e^{-\beta\epsilon_{P,L_{I}}}+\frac{[L_A]}{L_A}e^{-\beta\epsilon_{P,L_{A}}} \right)}{\frac{[P]}{P_D}\left(1+\frac{[L_I]}{L_I}e^{-\beta\epsilon_{P,L_{I}}}+\frac{[L_A]}{L_A}e^{-\beta\epsilon_{P,L_{A}}}\right)+1+\frac{[L_I]}{L_I}+\frac{[L_A]}{L_A}}.
\end{equation} 

Several additional factors influence gene expression \textit{in vivo}. First,
cAMP is synthesized endogenously by \textit{cya} and degraded by \textit{cpdA},
although both of these genes have been knocked out for the data set shown in
\fref[figSimpleActivationWT] (see Methods and Ref.~\cite{Kuhlman2007}).
Furthermore, cAMP is actively transported out of a cell leading to a smaller
concentration of intracellular cAMP. Following Kuhlman \textit{et al.}, we will
assume that the intracellular cAMP concentration is proportional to the
extracellular concentration, namely, $\gamma [M]$ (with $0 < \gamma < 1$)
\cite{Li2014b, Goldenbaum1979}. Hence, the concentration of active CRP satisfies
$\frac{[L_A]}{[L]} = p^L_{\text{act}}(\gamma [M])$ where the fraction of active
CRP $p^L_{\text{act}}$ is given by \fref[figCRPcAMPStatesWeights]\letter{A} as
\begin{equation} \label{eqPLactive}
p^L_{\text{act}}([M]) = \frac{\left(1+\frac{[M]}{M_L^A}\right)\left(1+\frac{[M]}{M_R^A}\right)}{\left(1+\frac{[M]}{M_L^A}\right)\left(1+\frac{[M]}{M_R^A}\right)+e^{-2\beta\epsilon}\left(1+\frac{[M]}{M_L^I}\right)\left(1+\frac{[M]}{M_R^I}\right)} \approx \frac{\frac{[M]}{\tilde{M}_L^A} \frac{[M]}{\tilde{M}_R^A}}{\frac{[M]}{\tilde{M}_L^A} \frac{[M]}{\tilde{M}_R^A}+\left(1+\frac{[M]}{M_L^I}\right)\left(1+\frac{[M]}{M_R^I}\right)}.
\end{equation}
In the last step, we have again introduced the effective dissociation constants
from \eref[eqMLAEffective][eqMRAEffective] and dropped any terms proportional to
$e^{-\beta \epsilon}$. In addition to these considerations, proteins \textit{in
	vivo} may experience crowding, additional forms of modification, and competition
by other promoters. However, since our primary goal is to understand how CRP
mutations will affect gene expression, we proceed with the simplest model and
neglect the effects of crowding, modification, and competition.

Because of the uncertainty in the dissociation constant $L_A$ between active CRP
and DNA (see Table~\ref{tableCRPDNABindingParamaters}), it is impossible to
unambiguously determine the transcription parameters from the single data set
for wild type CRP shown in \fref[figSimpleActivationWT]. Instead, we select one
possible set of parameters ($\frac{[P]}{P_D} = 130 \times 10^{-6}$,
$r_{\text{trans}} = 5 \times 10^5 \frac{\text{MU}}{\text{hr}}$, $\gamma = 0.1$,
$\epsilon_{P,L_{A}} = -3 \,k_B T$, and $\epsilon_{P,L_{I}} = 0 \,k_B T$) that is
consistent with the wild type data. Next, we inserted the other cAMP-CRP
dissociation constants (given in Table~\ref{tableCRPparamaters}) into
\eref[RNApBound] to predict the gene expression profiles of the CRP mutants.
\fref[figSimpleActivationWT] show the possible behavior of the
$\text{CRP}_\text{D/D}$ and $\text{CRP}_\text{WT/D}$ mutants. As expected,
replacing a WT subunit with a D subunit shifts the gene expression profile
leftwards since the D subunit has a higher cAMP affinity (see
\fref[figCRPcAMPligandTitration]\letter{A}). Interestingly, the substitution of
WT with D subunits comes with a concomitant increase in the maximum gene
expression because at saturating cAMP concentrations, a larger fraction of
$\text{CRP}_\text{D/D}$ is active compared to $\text{CRP}_\text{WT/WT}$ (96\%
and 68\%, respectively) as seen by using \eref[eqPLactive] and the parameters in
Table~\ref{tableCRPparamaters}. Note that we cannot predict the behavior of any
of the CRP mutants with S subunits due to the large uncertainty in
$\tilde{M}_\text{S}^A$.

\begin{figure}[t]
	\centering \includegraphics{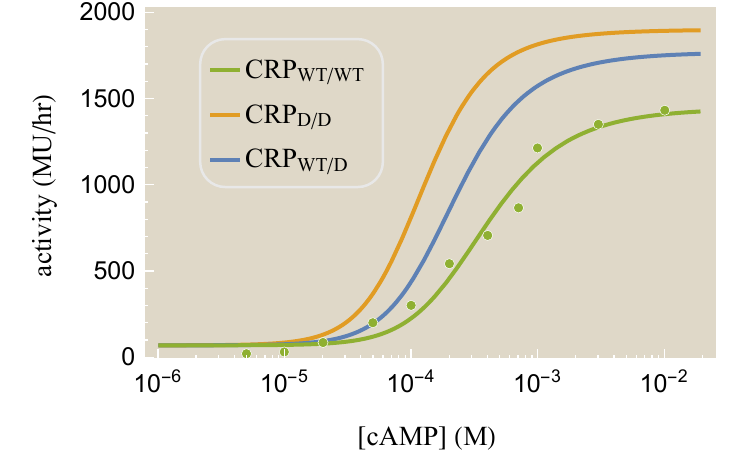}
	
	\caption{\textbf{Predicted gene expression profiles for a simple activation
		architecture.} Gene expression for wild type CRP (green dots from
	Ref.~\cite{Kuhlman2007}), where 1 Miller Unit (MU) represents a standardized
	amount of $\beta$-galactosidase activity. This data was used to determine the
	relevant parameters in \eref[RNApBound] for the promoter in the presence of
	$[L] = 1.5\,\mu\text{M}$ of CRP \cite{Cossart1985}. The predicted behavior of
	the CRP mutants is shown using their corresponding cAMP dissociation constants.
	Parameters used were $\frac{[P]}{P_D} = 130 \times 10^{-6}$, $r_{\text{trans}}
	= 5 \times 10^5 \frac{\text{MU}}{\text{hr}}$, $\gamma = 0.1$,
	$\epsilon_{P,L_{A}} = -3 \,k_B T$, $\epsilon_{P,L_{I}} = 0 \,k_B T$, and those
	shown in Tables~\ref{tableCRPparamaters} and
	\ref{tableCRPDNABindingParamaters}.} \label{figSimpleActivationWT}
\end{figure}

Lastly, we probe the full spectrum of phenotypes that could arise from the
activity function provided in \eref[RNApBound] for any CRP mutant by considering
all possible values of the cAMP-CRP dissociation constants $M_L^A$, $M_L^I$,
$M_R^A$, and $M_R^I$ in \eref[eqPLactive]. In particular, we relax our
assumption that cAMP binding promotes the CRP's active state, as a CRP mutation
may exist whose inactive state binds more tightly to cAMP than its active state.
\fref[figResponseTypes] demonstrates that given such a mutation, a variety of
novel phenotypes may arise. The standard sigmoidal activation response is
achieved when cAMP binding promotes the active state in both CRP subunits
($M_L^A < M_L^I$, $M_R^A < M_R^I$). A repression phenotype is achieved in the
opposite extreme when cAMP binding favors the inactive CRP state ($M_L^A >
M_L^I$, $M_R^A > M_R^I$). When one subunit is activated and the other is
repressed by cAMP ($M_L^A < M_L^I$, $M_R^A > M_R^I$ or $M_L^A > M_L^I$, $M_R^A <
M_R^I$), a peaked response can form. If the CRP subunits have the same affinity
for cAMP in the active and inactive states ($M_L^A = M_L^I$ or $M_R^A = M_R^I$),
then CRP will behave identically for all concentrations of CRP, generating a
flat-line response. It will be interesting to see whether these phenotypes can
be achieved experimentally.

\begin{figure}[t]
	\centering \includegraphics{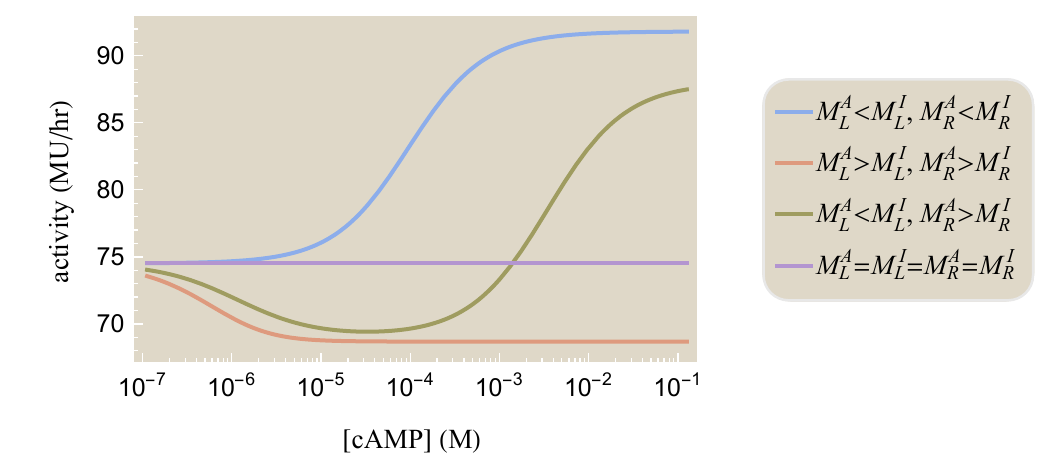}
	
	\caption{\textbf{The spectrum of input-output responses for mutant CRP in a
		simple activation architecture.} The possible gene expression profiles given by
	\eref[RNApBound] can be categorized based upon the cAMP-CRP binding affinity in
	each subunit. In all cases, we assumed $M_L^A = M_R^A = 3 \times
	10^{-6}\,\text{M}$. The activation response (blue) was generated using $M_L^I =
	M_R^I = 6 \times 10^{-6}\,\text{M}$. The repression response (orange) used
	$M_L^I = M_R^I = 10^{-7}\,\text{M}$. The peaked response (gold) used $M_L^I =
	10^{-7}\,\text{M}$ and $M_R^I = 300 \times 10^{-6}\,\text{M}$. The flat
	response used $M_L^I = M_R^I = 3 \times 10^{-6}\,\text{M}$. The remaining
	parameters were the same as in \fref[figSimpleActivationWT] together
	with $\epsilon = -3\,k_B T$.} \label{figResponseTypes}
\end{figure}

\section*{Conclusion}

The recent work of Lanfranco \textit{et al.}~provides a window into the
different facets of gene regulation through activation \cite{Lanfranco2017}.
Using insights from their \textit{in vitro} experiments, we can break down the
process of activation into its key steps, namely: (1) the binding of cAMP to
make the activator CRP competent to bind DNA (\fref[figCRPcAMPligandTitration]);
(2) the binding of CRP to DNA (\fref[figCRPDNAbindingMWCSimplest]); and (3) the
recruitment of RNAP to promote gene expression (\fref[figSimpleActivationWT]).
By concurrently modeling these processes, we begin to unravel relationships and
set strict limits for the binding energies and dissociation constants governing
these systems. One hurdle to precisely fixing these values for CRP has been that
many different sets of parameters produce the same degenerate responses (see
Supporting Information Section~A). This
parameter degeneracy is surprisingly common when modeling biological
systems \cite{Hines2014, Transtrum2015}, and we discuss how to account for it
within the MWC and KNF models of CRP. A key feature of our analysis is that it
permits us to identify the relevant parameter combinations for the system,
quantify how well we can infer their values, and suggest which future
experiments should be pursued to best constrain the behavior of the system.

Lanfranco \textit{et al.}~further explored how mutations in the cAMP binding
domain of one or both subunits of CRP would influence its behavior.
Specifically, they used three distinct subunits (WT, D, and S) to create the six
CRP mutants shown in \fref[figOverview]\letter{B}. We analyzed these constructs
using both the Monod-Wyman-Changeux and Koshland-Némethy-Filmer models of
molecular switching and demonstrated that either model can characterize all six
mutants using a self-consistent framework where each subunit is described by a
unique cAMP dissociation constant (see Table~\ref{tableCRPparamaters}). However,
these two models cannot be deemed successful simply because they yield curves
that can reproduce the data; the benefit of using mechanistic models involving
physical parameters is that the inferred values can be cross-checked against
other sources. For instance, the MWC model predicts that singly cAMP-bound CRP
will bind more tightly to DNA than unbound or doubly bound CRP, in line with the
structural knowledge of the system \cite{Heyduk1989, Pyles1996, Sharma2009}. And
while the KNF model can reproduce nearly identical CRP-DNA binding curves to
those generated by the MWC model, its inferred parameters imply that CRP should
bind equally well to its operator regardless of whether it is singly or doubly
bound to cAMP (see Table~\ref{tableCRPDNABindingParamaters}). Overall, these
results favor an MWC interpretation of the CRP system.

The models presented here suggest several avenues to further our understanding
of CRP. First, several groups have proposed that multiple CRP mutations (K52N,
T127, S128, G141K, G141Q, A144T, L148K, H159L from Refs.~\cite{Lin2002,
	Youn2006, Youn2008}) only affect the free energy difference $\epsilon$ between
the CRP subunit's active and inactive states while leaving the cAMP-CRP
dissociation constants unchanged. It would be interesting to test the framework
developed here across CRP mutants specifically designed to vary these
parameters, since the $\epsilon$ dependence of the system is completely
relegated to the effective dissociation constants (see
\eref[eqMLAEffective][eqMRAEffective]).

Second, the MWC and KNF models can be used to predict how the CRP mutants
generated by Lanfranco \textit{et al.}~would behave \textit{in vivo}. We
calibrated the $\text{CRP}_\text{WT/WT}$ gene expression profile using data from
Ref.~\cite{Kuhlman2007} and suggested how the remaining CRP mutants may function
within a simple activation regulatory architecture given the currently available
data (see \fref[figSimpleActivationWT]). It would be interesting to measure such
constructs within the cell and test the intersection of our \textit{in vivo} and
\textit{in vitro} understanding both in the realm of the multi-step binding
events of CRP as well as in quantifying the effects of mutations.

Finally, we note that both the MWC and KNF models can serve as a springboard for
more complex descriptions of CRP. It has been debated whether the first cAMP
binding event inhibits a second cAMP from binding or if doubly bound CRP has a
weaker affinity to DNA \cite{Liu2017, Yu2012, Heyduk1990, Takahashi1989,
	Heyduk1989}, and such modifications are straightforward to add to the models
discussed above. However, a key advantage of the simple frameworks presented
here lies in their ability to \textit{predict} how different CRP subunits
combine. For example, in Supplementary Information
Section~B we demonstrate how the data from the three
symmetric CRP mutants in \fref[figCRPcAMPligandTitration]\letter{A} can be used
to characterize the asymmetric mutants in
\fref[figCRPcAMPligandTitration]\letter{B}. It would be interesting to see
whether such predictions continue to hold as more mutant subunits are
characterized. Such a framework has the potential to harness the combinatorial
complexity of oligomeric proteins and presents a possible step towards
systematically probing the space of mutations.

\section*{Methods}

As described in Ref.~\cite{Lanfranco2017}, the fractional CRP occupancy data in
\fref[figCRPcAMPligandTitration] was measured \textit{in vitro} using
8-anilino-1-naphthalenesulfonic acid (ANS) fluorescence which is triggered by
the conformational change of cAMP binding to CRP. The CRP-DNA anisotropy data in
\fref[figCRPDNAbindingMWCSimplest] was measured \textit{in vitro} by tagging the
end of a $32\,\text{bp}$ \textit{lac} promoter with a fluorescein molecule and
measuring its anisotropy with a spectrometer. When CRP is bound to DNA,
anisotropy arises from two sources: the fast bending of the flanking DNA
sequence and the slower rotation of the CRP-DNA complex. Sources of error
include oligomerization of CRP, the bending of the flanking DNA, and nonspecific
binding of CRP to the DNA.

The \textit{in vivo} gene expression data was taken from Kuhlman et al.~using
the \textit{lac} operon \textit{E. coli} strain TK310 \cite{Kuhlman2007}. This
strain had two genes knocked out: \textit{cya} (a gene encoding adenylate
cyclase, which endogenously synthesizes cAMP) and \textit{cpdA} (encoding
cAMP-phosphodiesterase, which degrades cAMP within the cell). Experiments were
done at saturating concentrations of inducer ($[\text{IPTG}] = 1\,\text{mM}$) so
that Lac repressor negligibly binds to the operator. In this limit, the only
transcription factor affecting gene expression is the activator CRP. Gene
expression was measured using $\beta$-galactosidase activity.

\section*{Supporting Material}

Supporting Materials with the aforementioned derivations are available online
together with a Mathematica notebook that contains all the data, reproduces the
fitting (using both nonlinear regression and MCMC), and generates the plots from
the paper.

\section*{Author Contributions}

TE, JD, and RP performed the research. TE and RP wrote the manuscript.

\section*{Acknowledgements}

We thank Lacramioara Bintu for bringing the recent developments on CRP to our
attention as well as Terry Hwa, Tom Kuhlman, and Michael Manhart for helpful
discussions. All plots were made entirely in \textit{Mathematica} using the
CustomTicks package \cite{Caprio2005} with data obtained from the authors or
using WebPlotDigitizer \cite{Rohatgi2017}. This work was supported in the RP
group by La Fondation Pierre-Gilles de Gennes, the Rosen Center at Caltech, and
the National Institutes of Health through DP1 OD000217 (Director's Pioneer
Award), R01 GM085286, and 1R35 GM118043-01 (MIRA). We are grateful to the
Burroughs-Wellcome Fund for its support of the Physiology Course at the Marine
Biological Laboratory, where part of the work on this work was done, and for a
post-course research grant (JD).


%% file: 2.1_Text_References.tex

\section*{Supporting Citations}

References \cite{Marzen2013, Rodrigues2016, Auerbach2012} appear in the Supporting Material.